\documentclass{ws-ijmpcs}

\begin{document}

\markboth{F. G. de Oliveira, R. M. Marinho Jr., J. G. Coelho and N. S. Magalhaes}
{Data Analysis of Gravitational Waves Signals from Millisecond Pulsars}

%
\catchline{}{}{}{}{}
%

\title{Data Analysis of Gravitational Wave Signals from Millisecond Pulsars
}

\author{Fernanda G. Oliveira, Rubens M. Marinho Jr and Jaziel G. Coelho}

\address{Instituto Tecnol\'{o}gico de Aeron\'{a}utica, Pra\c{c}a Marechal
Eduardo Gomes 50, S\~{a}o Jos\'e dos Campos,  SP 12228-900, Brazil}

\author{Nadja S. Magalhaes}

\address{Universidade Federal de S\~{a}o Paulo, DCET, Rua S\~{a}o Nicolau 210, Diadema, SP 09913-030, Brazil}

\maketitle

\begin{history}
\received{Day Month Year}
\revised{Day Month Year}
\end{history}

\begin{abstract}
The present work is devoted to the detection of monochromatic gravitational wave
signals emitted by pulsars using ALLEGRO's data detector. 
We will present the region (in frequency) of millisecond pulsars 
of the globular cluster 47 Tucanae (NGC 104) in the band of detector. 
With this result it was possible to analyse the data in the frequency ranges of the pulsars J1748-2446L and J1342+2822c, searching for annual Doppler variations using power spectrum estimates for the year 1999. We tested this method injecting a
simulated signal in real data and we were able to detect it. 

\end{abstract}

\ccode{PACS numbers: 11.25.Hf, 123.1K}

\section{Introduction}	

The focus of this work was to analyse ALLEGRO's data$^[$\cite{aquisition}$^]$ for the year 1999 taking 
into account the effect due to the orbital motion of the Earth for
 specific frequencies, $891.0$ Hz and $923.4$ Hz, that correspond to the pulsars located in 47 Tucanae (NGC 104) named 1748-2446L
and J1342+2822c, respectively. This analysis was based on estimates of power spectrum 
of the data using averaged modified periodograms$^[$\cite{iwara10}$^]$ which 
reinforce the presence of peaks due to monochromatic signals. 

\section{The Characteristic Amplitude of a Pulsar's Gravitational Waves}
\begin{figure}[ht]
\begin{center}
    \fbox{\includegraphics[width=8cm]{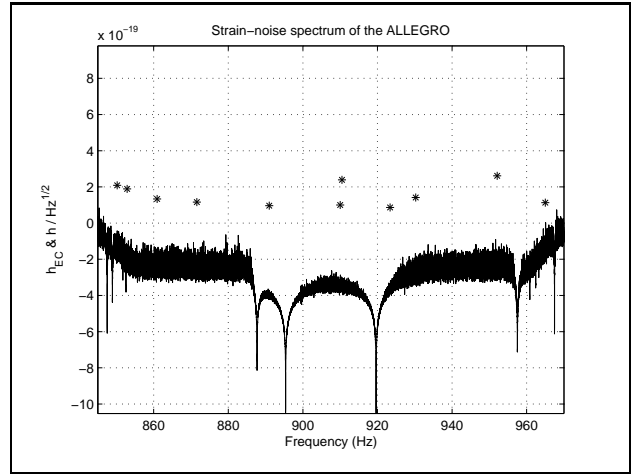}}
    \caption{The figure shows the gravitational strain $h_{EC}$ (EC stands for Energy Conservation or spin-down limit) for the
known pulsars in 47 Tucanae (NGC 104). The strain $h_{EC}$ is compared
to the strain-noise spectrum of the detector ALLEGRO (in units of $h/\sqrt{ \rm Hz}$).}
    \label{strain}
\end{center}
\end{figure}

Pulsars with non-axisymmetric rotation are expected to emit monochromatic gravitational wave signals (MGW). 
The amplitude of gravitational waves (GW) emitted by a rotating neutron star (NS) can be expressed 
in terms of the NS rotation period $P$, the distance to the Earth $r$, the moment of inertia $I$ and the
ellipticity $\epsilon$ resulting from the distortion process as$^[$\cite{gourgoulhon}$^]$:
\begin{equation}
 h_c= 4.21 \times 10^{-4}\left(\frac{\rm ms}{P}\right)^2\left(\frac{\rm kpc}{r}\right)\left(\frac{I}{10^{38}\rm kgm^2}\right)^2 \left(\frac{\epsilon}{10^{-6}}\right).
\label{hc}
\end{equation}
Its value depends on the physical mechanism that makes the star
non-axisymmetric and is highly uncertain. 
The values of $h_c$ resulting from Eq. (\ref{hc}) are show in the Figure \ref{strain} for 
millisecond pulsars in 47 Tucanae.

\section{Determination of the Observation Time}
In the present analysis we are interested only in the annual Doppler shift that a monochromatic, continuous gravitational wave signal should experience, so we need to choose an observation time, $\Delta t $, such that the diurnal Doppler shift, $\Delta\nu_d$,
remains in the same frequency bin $\Delta f=1/\Delta t$. The minimum size of this bin, $\Delta f_{min}$, corresponds to the 
maximum diurnal Doppler shift, $\Delta\nu_{dmax}=\Delta f_{min}=1/\Delta t_{max}$.
The maximum diurnal Doppler shift happens when the Earth and the star are in the line of the nodes\cite{lightcone}:
\begin{equation}
\Delta\nu_{dmax} = \nu_s\frac{2wr}{c},
\end{equation}
where $w$ and $r$ are the angular velocity of rotation and the radius of the Earth, respectively.
The annual Doppler shift in a full year of observation is,
\begin{equation}
    \Delta\nu_{a} = \nu_s\frac{2R\Omega}{c},
\end{equation}
where $R$ is the average radius of the Earth's orbit around the Sun and $\Omega$ is its angular velocity in this orbital motion.
The values of the Doppler shifts obtained for the pulsars radiating with frequency $\nu_s$ are given in Table 1.
\begin{table}[ht]
\begin{center}
\title{Table 1: Doppler shifts and observation times for two pulsars in 47 Tucanae expected to emit gravitational waves in the frequencies $\nu _s$.}
\begin{tabular}{l c c c c c c c}  \hline \hline
 \hspace{1cm}Pulsar &  $\nu_s$  & \ $\Delta\nu_{dmax}$   & \   & $\Delta\nu_{a}$ & \   &     $\Delta t_{max}$                \\ \hline
 PSR J1748-2446L &  891.0 Hz      & \ 2.8 mHz  & \   & $ 0.1770$ Hz         & \   &     362 s            \\ \hline
 PSR J1342+2822c &  923.4 Hz    & \ 2.9 mHz   & \   & $0.1835$ Hz          & \   &     350 s        \\ \hline \hline 
\label{tabela}
\end{tabular}
\end{center}
\end{table} 

In order to eliminate the daily Doppler shift we used $\Delta t=300$ s for the observation time in our data analysis.

\section{The Strain-Noise Sensitivity of ALLEGRO}
In Figure \ref{strain} we present the region of MGW signals from pulsars in the strain-noise spectrum of ALLEGRO. 
This figure shows the gravitational strain $h_{EC}$ for the
known pulsars in the band of ALLEGRO from the ATNF catalog. 
This quantity was derived
assuming that all observed spin-down is due to energy loss
caused by emission of gravitational radiation (and no other
braking mechanisms$^[$\cite{santostasi}$^]$). The strain for the pulsars is compared
to the noise sensitivity curve (in units of $h/\sqrt{Hz}$) for the
ALLEGRO detector. 

\begin{figure}[ht]
\begin{center}
    \fbox{\includegraphics[width=8cm]{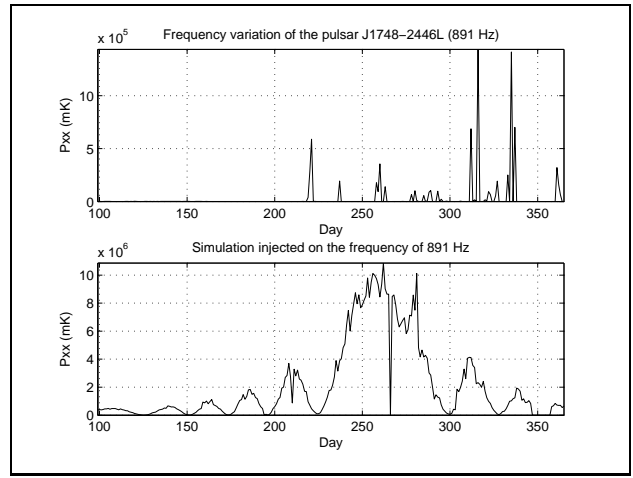}}
    \caption{Upper plot: Variation of the power spectrum for PSR J1748-2446L. Lower plot: Variation of the power spectrum for the simulated signal added to the data.}
    \label{J1748}
\end{center}
\end{figure}

\begin{figure}[ht]
\begin{center}
    \fbox{\includegraphics[width=8cm]{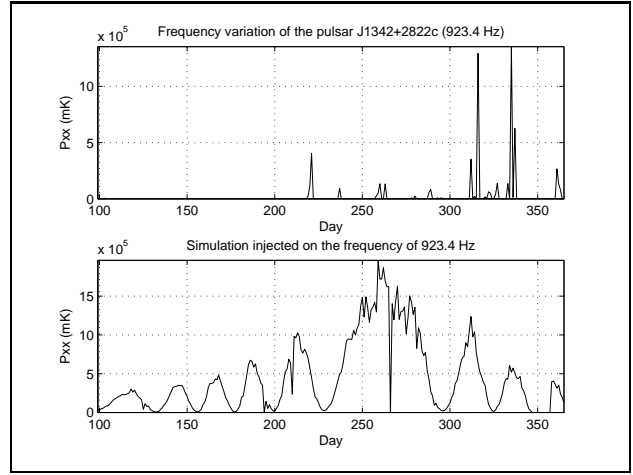}}
    \caption{Upper plot: Variation of the power spectrum for PSR J1342+2822c. 
Lower plot: Variation of the power spectrum for the simulated 
signal added to the data.}
    \label{J1342}
\end{center}
\end{figure}

\section{The Data Analysis}
The goal of the analysis was to search for an annual Doppler shift in ALLEGRO's data. 
For this we stablished an observation time $\Delta t = 300$ s so that the frequency of a possible observed signal would not change 
from one bin to another during the day. We have taken the power spectral density for 266 days of the year 1999.
We fixed our attention on the bins that contained the frequencies 891.0 Hz and 923.4 Hz 
that would correspond to GW radiated respectively by the pulsars J1748-2446L and J1342+2822c,
looking for an excess energy in these bins during those 266 days.
We have chosen these two pulsars because their radiated frequencies were
near the  frequencies where the detector is the most sensitive$^[$\cite{aquisition}$^]$.
We simulated a GW  signal with dimensionless amplitude $ h = 2.6\times 10^{-17}$
and added it to ALLEGRO's data. 
The results of this analysis are shown in Figures \ref{J1748} and \ref{J1342}.

\section{Conclusions}

In this analysis we are not able to identify any Doppler modulation in real data, as seen from the upper plots in Figures \ref{J1748} and \ref{J1342}. However, it was possible to test our data analysis procedure for detection of monochromatic
gravitational wave signals since we were able to notice the simulated signal buried
in the noise (lower plots in Figures \ref{J1748} and \ref{J1342}). The calculation of the detection probability using the Neyman-Pearson criterion will be the subject of a forthcoming paper.

\end{document}